# Using Humanoid Robot to Instruct and Evaluate Performance of a Physical Task

Shawn N. Gieser, Joseph Tompkins, Ali Sharifara, Fillia Makedon

HERACLEIA Human Centered Computing Laboratory
Department of Computer Science and Engineering
University of Texas at Arlington, Arlington, TX 76010

{shawn.gieser, ali.sharifara, makedon}@uta.edu, joseph.tompkins@mavs.uta.edu

*Abstract*—In this paper, we present a tool to assess a user's ability to change tasks. To do this, we use a variation of the Box and Blocks Test. In this version, a humanoid robot instructs a user to perform a task involving the movement of certain colored blocks. The robot changes randomly change the color of blocks that the user is supposed to move. Canny Edge Detection and Hough Transformation are used to assess user perform the robot's built in camera. This will allow the robot to inform the user and keep a log of their progress. We present this method for monitoring user progress by describing how the moved blocks are detected. We also present the results from a pilot study where users used this system to perform the task. Preliminary results show that users do not perform differently when the task is changed in this scenario.

## I. INTRODUCTION

As social low-cost robots continue to be developed, their adaption for educational purposes has been increased, especially for children with disabilities or users in the aged population. These Humanoid Robots have been especially useful in helping users with Autism Spectrum Disorder to connect and establish a relationship [11]. This brings up the question of how everyday users would interact with robots. How would everyday users with no special needs respond to robots issuing instructions and evaluating them?

In this paper, we present a prototype system to assess user performance while performing tasks while being instructed by a humanoid robot. This system will detect how many blocks a user moves during a modified version of the Box and Blocks Test, as well as monitor the rate of speed at which the user performs the tasks. We will also present the results of a pilot study, which shows the accuracy of our block detection technique and compares times that users took during these tasks. We will first cover work of others related to our system. Then we will describe our system, including the equipment used, the block detection technique, and our experimental setup and procedure. We will then present and discuss our results. Lastly, we will present our conclusions and go over future extensions of development and testing.

## II. RELATED WORK

Socially assistive robots are becoming increasingly popular, especially when it comes to being able to interact with humans and assist them to enhance skill-based tasks [3]. These kinds of robots aim to help humans via social interaction, which instructs, observes, and gives feedback. In the following, several methods which have used socially assistive robots are described.

Litoiu and Scassellati [4] proposed a method to instruct physical activities using humanoid robot. The authors concentrated on identifying problems with movements and prioritizing them. For classification purposes, a supervised learning method was used to classify the problems to work on, and recommend the design of several user studies to determine the effectiveness of the algorithm.

Similarly, Yamada and Miura [14] proposed an approach for human to robot teaching. The authors claimed that their proposed robot advisor instructs a human learner how to perform a block assembly task. The authors believe that the key to understanding an interaction is to properly determine basic probability assignments, which are set manually based on the degree of knowledge transfer for each task and instruction.

Likewise, Shen and Wu [13] evaluate the user experience of robots for elderly people. Their results show that a robot was more effective and preferred by subjects over a human instructor for delivering information and instructing physical exercise.

Sauppe and Mutlu present an autonomous instructional robot. The authors compare different instructional strategies, which can have an effect on user performance and experience. Their analysis of human instructor interactions identified two key instructional strategies, including grouping instructions and summarizing the result of subsequent instructions. They implemented the above strategies by using a humanoid robot which is able to correct its mistakes and misunderstandings. The authors also claimed that their findings are outstanding for the design of instructional robots. [10].

Schneider et al. proposed a framework for coordinating motivational interaction scenarios with robots in the instruction of physical activities. The authors have tested their framework with three different physical activity models where they have used the same motivational interaction scenario. The authors concluded that their model can be applied to systematically test different aspects of motivation using a robot in coaching physical activities [12].

In the same manner, Ding and Chang [2] developed a Kinect-sensor-based sport instructor robot system for rehabilitation and physical activity training of the aged population. They used Kinect camera in order to detect and recognize the gestures of people. The Kinect is used to check the correctness of the active gesture. The authors have claimed that the combination of humanoid robot and Kinect sensor worked successfully in order to perform gesture recognition. They have tested their framework with three different exercise activities.

Park and Kwon [8] recently published a scientific research about feasibility of teaching children with cognitive disabilities from robot instructors. Moreover, they have concluded that their results proved that children who were trained with a robot instructor notably enhanced their skills and functional knowledge, especially when children repeated the training session to get better results.

As can be inferred from the related work, robot instruction methods show outstanding results in comparison with human instructor's due to several advantages which robots have over humans including: 1. Robots do not get tired, 2. Robots are not going to judge the user, 3. Robots are able to repeat their instruction an unlimited number of times without getting tired.

III. EXPERIMENTAL SETUP AND PROCEDURE

This section will detail the setup of our system and experiment, our methods for detecting blocks, and lastly, the procedures used to collect data and verify our system.

*A. Experimental Setup*

The task we asked users to perform was a modified version of the Box and Blocks Test [6]. In this version, there were only seventy-three blocks instead of one-hundred fifty. There were three different colors of blocks: thirty red, thirty green, and thirteen blue. There was also no time constraint of one minute. The task lasted as long as necessary for the user to complete it. All other rules, such as only picking up one block at a time and that the user's hand has to completely cross the partition of the box when moving a block, were still enforced. Figure 1 shows the box and blocks test used for the experiment.

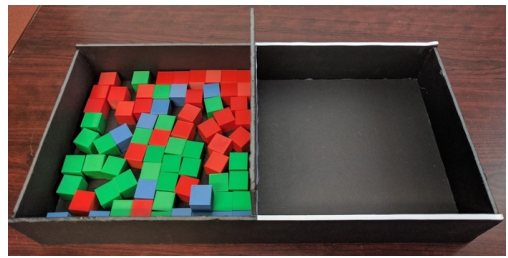

Fig. 1: The Box and Blocks Test used for this task.

The Nao Humanoid Robot [9], Figure 2, was used to give instructions to user while performing the task. The Nao Robot told users what color block to move. This was used in assessing a user's abilities to switch tasks. This was accomplished by having the Nao Robot's head pointed towards the side of the box where the user was moving blocks. After each block was moved, the user tapped the Nao Robot's head so that progress could be recorded. The Nao Robot would then use these progress points to assess the user's ability to move the correct blocks, as well as see how long it took to move each block based on the time-stamp of when the head was tapped. Figure 3 shows the setup of the Box and Blocks and the Nao Robot.

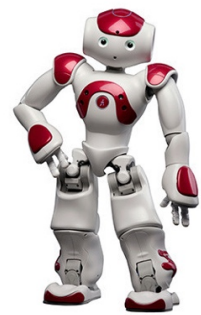

Fig. 2: Aldebaran's NAO robot - autonomous, programmable humanoid robot.

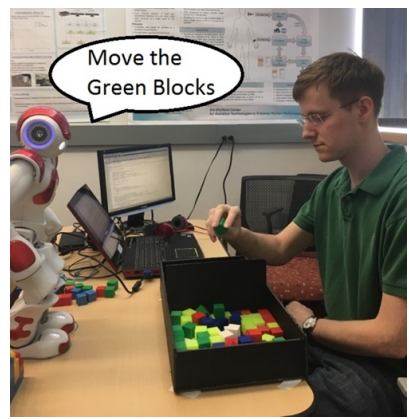

Fig. 3: Experimental Setup of the Nao and Box and Blocks Test

*B. Computer Vision for Block Detection*

For the purposes of this system, Canny edge detection [1] and Hough transforms [5] were heavily relied upon in the implementation of the computer vision algorithms. The resulting lines produced by a Hough transform are used to detect geometric features indicating that the top of a block has been found. Namely, the algorithm determines which lines might produce squares within a specified tolerance threshold. This method was chosen in order to minimize the number of necessary constraints established on how users may place blocks, and thus further preserve the integrity of the original Box and Blocks test.

Object detection is divided into to two main stages, each split into two sub-stages. The main stages are comprised of Image Area Reduction and Block Identification. Once these processes are complete, the location of the blocks is known, and blocks are counted based on their RGB color space information.

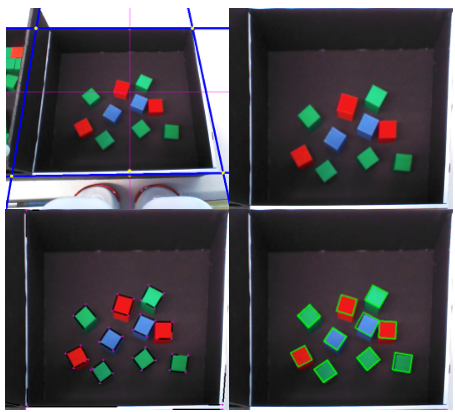

Fig. 4: Block detection process

Detection of the box is the first sub-stage of the Image Area Reduction stage. The target area is the portion of the box where users will be placing the blocks. The resulting image produced by the Canny edge detection algorithm provided in the OpenCV [7] library is then processed by the probabilistic Hough transform algorithm provided by the same library. This result is then used to determine the perimeter of the desired area, as shown in the top left portion of Figure 4. Specifically, the lines produced by the Hough transform are used to determine the corners of the target area of the box. For the second sub-stage of Image Area Reduction, a perspective transformation is performed based on the corners produced by the box detection process. The result is an image which attempts to mimic a top-down view of the blocks. This is to provide an image of the blocks with clearly defined edges for the next main stage. The resulting image is shown in the top right of Figure 4.

The first sub-stage of the Block Identification stage is detecting the edges and corners of the blocks. This process utilizes, once again, the Canny edge and probabilistic Hough transform algorithms provided in the OpenCV library. The output resulting from the Hough transform is then processed in order to filter out line segments that do not meet certain criteria. The criteria focus on reducing the amount of resulting lines to consider only those that could potentially form squares, indicating the top of a block has been found. The first criterion that line segments must meet is that they intersect at a right angle at least once with another line. These line segments then must have end-points within a specified distance of an intersection point. From those remaining, line segments that are within a specified distance from other line segments that they intersect with are accepted. The resulting line segments and intersection points are shown in the bottom left of Figure 4.

The final stage of block detection uses the cumulated result from this process to determine the tops of the squares. This is accomplished by iterating through a clockwise or counter-clockwise path of correlated intersection points. See the bottom right of Figure 4 for the final result.

Color detection was performed using RGB color space image, as seen in Figure 4. As of this version of the system, color detection was reduced to the trivial matter of finding the maximum RGB color value of a pixel in the center of a detected block. This is because the system was constrained to using only red, green, and blue blocks for simplicity. Color detection could be expanded in the future by relying on other methods to determine color, such as HSV color space masking.

By reducing the initial image area to the target area and transforming the result to mimic a top-down perspective, the problem of block detection becomes one of detecting simple geometric shapes. The squares, representing the top of a block, were able to be determined using the line information resulting from performing a probabilistic Hough transform on the perspective corrected image. Figure 5 shows the system architecture. In this, you can see how the user and robot interact and how the robot performs the error analysis using the computer vision methods described above.

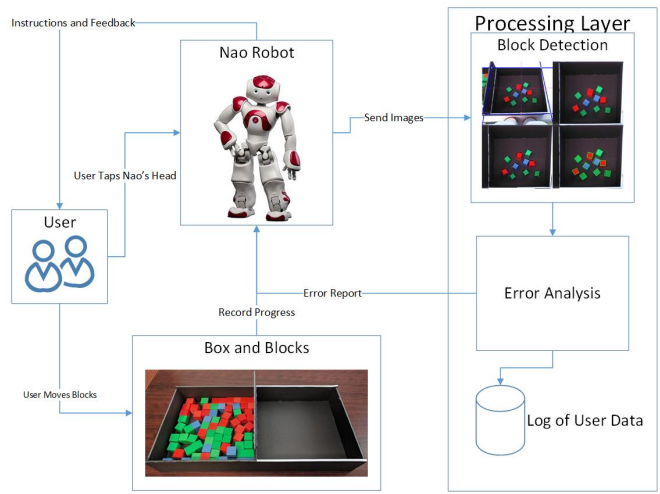

Fig. 5: System Architecture Schema.

## C. Procedures

This procedure received approval from The University of Texas at Arlington's Institutional Review Board, reference number 2017-0557. After obtaining consent, users were placed at a table in front of the Box and Blocks Test and the Nao Robot.

The test administrator (admin) positioned the box so that the test can be administer on the user's dominant hand. The admin started the program once the user was in position. The Nao robot then informed the user of the task and asked the user to tap its head when the user is ready to begin.

Once the user told the robot he was ready, the Nao randomly determines a color from the colors Red, Green, and Blue. The color chosen was told to the user. The Nao then randomly decided how many blocks of that color the user had to move on a range from two to three. The Nao instructed the user to move a block. The user moved a block and tapped the Nao's head to signify that the he is ready for the next instruction. The Nao recorded the user's progress and sent the data to a computer for processing. The Nao kept asking the user to move blocks until the user has moved the proper number of blocks for that color. This is called a cycle.

Once a cycle is complete, the Nao determined a new color and began a new cycle. Cycles were repeated until a total of five cycles were completed. Once the cycles were completed, the Nao asked the user to switch hands. The admin repositioned the box so that the user can perform the task with their non-dominant hand. The user then completed another five cycles.

Once both hands were completed, the Nao received the results of the processed images in the form of the number or errors detected. The Nao gave feedback to the user by informing him about the number of errors committed during the task. Once the feedback was given, the user was told he could leave and the admin reset the task for the next user.

**Algorithm 1** shows the procedure in pseudo-code from the point of view of the Nao Robot and Error Analysis Program.

| **Algorithm 1** Pseudocode for how the robot instructs, interacts, observes, and provides feedback to the user |
|---|
| 1: procedure BOXAndBlocksWithNaoRobot |
| 2:     $NumberofErrors \leftarrow 0$ |
| 3:     $NumberofBlocksMOved \leftarrow 0$ |
| 4:     $NumberCyclesComplted \leftarrow 0$ |
| 5:     $ColorToMove \leftarrow null$ |
| 6:     $NumberToMove \leftarrow 0$ |
| 7:     $HandDone \leftarrow 0$ |
| 8: Wait for user: |
| 9:     Wait for user to tap head when user is ready |
| 10: Color Decision: |
| 11:     $ColorToMove \leftarrow Random\ (Red, Green, Blue)$ |
| 12:     Instruct user to move ColorToMove Blocks |
| 13: Number of Decision: |
| 14:     $NumberToMove = Random(2,3)$ |
| 15: Move Block |
| 16:     Instruct user to move a block |
| 17:     Wait for user to move block and tap head |
| 18:     Record user progress |
| 19:     $NumberBlocksMoved \leftarrow NumberBlocksMoved + 1$ |
| 20: **if** $NumberBlocksMOved\ != NumberToMove$ **then** |
| 21:     **goto** *Move Block* |
| 22:     $NumberCyclesCompleted \leftarrow NumberCyclesCompleted + 1$ |
| 23: **if** $NumberCyclesCompleted\ != 5$ **then** |
| 24:     $NumberBlocksMoved \leftarrow 0$ |
| 25:      **goto** Color Decision |
| 26:     $HandsDone \leftarrow HandsDone + 1$ |
| 27: **if** $HandsDone == 1$ **Then** |
| 28:     $NumberBlocksMoved \leftarrow 0$ |
| 29:     $NumberCyclesCompleted \leftarrow 0$ |
| 30:     **goto** *Wait for User* |
| 31:     *Calculate errors using Computer Vision* |
| 32:     *state number of errors to user* |
| 33: *End* |

## IV. RESULTS & DISCUSSION

We had six users perform this task. All users were of good health with no physical injuries or mental impairments. Each person followed the same procedure. A total of 149 blocks were moved, giving us the same number samples of user progress.

Table I shows the accuracy of our error detection technique. Actual errors are instances when the user made an error during the task. Perceived errors are instances when our system detected an error. Our method of detecting blocks is not perfect, since it yielded an accuracy of 65.10% when assessing human motion. The other 34.90% of the time, our method detected an error that was not made by the user.

Table I:
Block Detection Error Results and Accuracy

| User Number | No. of Blocks Moved | Actual Errors | Perceived Errors | Accuracy |
|---|---|---|---|---|
| 1 | 22 | 0 | 9 | 59.09% |
| 2 | 23 | 0 | 4 | 82.61% |
| 3 | 25 | 0 | 7 | 72% |
| 4 | 26 | 0 | 8 | 69.23% |
| 5 | 28 | 3 | 18 | 46.43% |
| 6 | 25 | 0 | 9 | 64% |
| Total | 149 | 3 | 55 | 65.10% |

There were six different types of errors that occurred in our block detection method. The first was that the system would detect an extra block, which can be seen in the top-left of Figure 6. When this error materializes, the square representing the extra block usually differs in size from the average block. Eliminating those squares that deviate from the average should prevent this error from materializing.

The second type of error was the system would not detect a block, which can be seen in the top-center of Figure 6. This occurs when the computer vision process is not able to identify enough edges of the top of a block to determine it is presence, either due to occlusion, lighting, or some other factor preventing the Hough transform from detecting an edge. Further image processing prior to the computer vision process should be able to correct the issue and bring the rate of this particular error to an acceptable level.

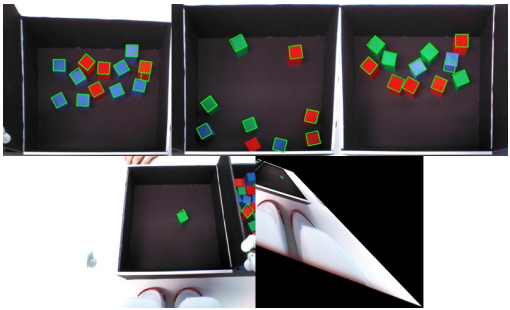

Fig. 6: Example of Errors with Block Detection. Top-Left: Detecting an extra block. Top-Center: Not detecting a block. Top-Right: Misaligned block detection. Bottom-Left: Nao Seeing the other Side. Bottom-Right: Error in perspective

The third type of error was the system would misalign a block, which can be seen in the top-right of Figure 6. This occurs when the computer vision process misinterprets the bottom edges of the sides of a block as belonging to the top of the block. This error rarely results in a false positive, because RGB color information is still present at the center of the square. Despite this, two proposed solutions to correct this error would be the use of HSV color space to discriminate between the top and sides of a block, and/or the use of stereoscopic vision included in the Nao robot's hardware features to collect depth-of-field information.

The fourth type of error was the system would detect the other side, which can be seen in the bottom-left of Figure 6. Occasionally a failure to produce all four edges of the target area of the box occurs. Block detection will resume despite not having an image that has been perspective corrected, resulting in blocks on the other side of the divider being detected. A possible solution would be to modify the box further by adding computer vision landmarks to aide image processing. The fifth type of error was the system would not properly perform perspective correction, which can be seen in the bottom-right of Figure 6. Similar to the fourth type of error, the referenced image follows a failure of the box detection process to identify all four edges of the target area. The perspective correction transformation requires all four corners of the target area to be known in order to correctly transform the image. The proposed solution for the fourth type of error could also be used to resolve this error.

The last type of error was the system would count previous mistakes. The system does not currently track unique errors made by the user. This means that if the user places a single incorrect block in one instance or cycle, that error will persist, and therefore be counted, in all of the following instances.

Next, we will look at the speed at which the users completed the task. Table II shows the average speed it took users to move a block both when there was a color and when there was not. Two users averaged faster times when there was a color change compared to when there was not. Three users performed faster on average when there was no color change. One user had very similar average times between the two different scenarios. This can lead us to believe that there is no difference between the two difference scenarios. This claim, however, cannot be fully supported due to a small number of users.

As far as being able to assess the user's performance when changing tasks, the current accuracy rate for detecting blocks makes this very difficult. We can see that User Five had significantly more perceived errors than the rest of the users. This is promising since this user was the only user to make real errors. We can then assume that this user had a harder time switching between different colors. The times show that Users Three and Six and significant more time taken (over one second) to switch colors compared to that of User Two (under one second). We could then assume that those two users also had a harder time switching tasks but were able to successfully do so.

There were a couple limitations to this pilot study. This first, which was previously stated was the small number of users. This may not only have affected the average speed, but may also have affected the number of errors committed during the task. Having a larger user base will increase the likelihood of having errors and a greater variation of speed of task performance. Secondly, the design of the study only provided one mode of instruction. This limited the possible results of this study. If multiple forms of instruction were included, then comparison could be made between the number of errors and speed between the different modes of instruction.

## V. CONCLUSIONS

In this paper, we presented our prototype for providing instructions to a user for a task from a humanoid robot. We also used this robot to assess user performance by determining the number of errors and timing how long it takes the user to perform each instance of the task. We also show the results of a very limited pilot study to evaluate our block and error detection methods and to see how long it takes users to perform a task when part of that task changes. Our technique for block and error detection can be improved upon to yield a higher accuracy. Even though our study limited the strength of our results, preliminary results do show that there is not a substantial difference between a user's ability to perform tasks when there is a change. This may or may not be caused by using the robot to provide the instructions. Further studies would be needed to determine whether this is true.

## VI. FUTURE WORK

There are many different ways to extend this pilot study. The first few ways deal with the limitations of the study previously mentioned. The first is that improve the block detection method to reduce the number of errors. The ways to do this have been detailed in the Results and Discussion Section. This will improve our system to allow better assessment of users. Second, increase the number of users involved in the study. This will allow for further verification of our block detection method and to develop a more formal assessment of user performance. Third, we would compare results of using a robot to provide instructions to that of other methods. These methods could be a human giving instructions, or a color being displayed on a monitor screen. This will allow us to validate if a robot providing the instructions would be better than other methods of instruction.

Lastly, we would improve our system by adding other modalities to assess user performance. We could use a low-cost EEG headset to get the user's attention level. This would allow us to monitor the user's attention level during task changes and when the user makes an error. We could also attach an accelerometer to the user's arm. Knowing the motion of the user's arm would enable the detection of corrections. The user could pick up a block, start moving it to the other side, realize it is the wrong color, and have to move back and pick up a block of the proper color. By tracking the user's arm motion, we can detect if a user has to go back and pick up another block. Adding these two modalities could greatly aid in our ability to assess user performance.

## VII. ACKNOWLEDGMENT

We would like to extend our thank the volunteers that participated in this study. Any opinions, findings, and conclusions or recommendations expressed in this publication are those of the author(s) and do not necessarily reflect the views of the National Science Foundation.

## REFERENCES

[1] John Canny. A computational approach to edge detection. IEEE Transactions on pattern analysis and machine intelligence, (6):679–698, 1986.

[2] IJ Ding and Yu-Jui Chang. On the use of kinect sensors to design a sport instructor robot for rehabilitation and exercise training of the elderly. Sensors and Materials, 28(5):463–476, 2016.

[3] Daniel Leyzberg, Samuel Spaulding, and Brian Scassellati. Personalizing robot tutors to individuals' learning differences. In Proceedings of the 2014 ACM/IEEE International Conference on Human-robot Interaction, HRI '14, pages 423–430, New York, NY, USA, 2014. ACM.

[4] Alexandru Litoiu and Brian Scassellati. Personalized instruction of physical skills with a social robot. Machine Learning for Interactive Systems at AAAI, 2014.

[5] Jiri Matas, Charles Galambos, and Josef Kittler. Robust detection of lines using the progressive probabilistic hough transform. Computer Vision and Image Understanding, 78(1):119–137, 2000.

[6] Virgil Mathiowetz, Gloria Volland, Nancy Kashman, and Karen Weber. Adult norms for the box and block test of manual dexterity. American Journal of Occupational Therapy, 39(6):386–391, 1985.

[7] OpenCV. Opencv library. http://opencv.org/.

[8] Eunil Park, Sang Jib Kwon, et al. I can teach them: The ability of robot instructors to cognitive disabled children. Journal of Psychological and Educational Research (JPER), 24(1):101–114, 2016.

[9] SoftBank Robotics. Discover nao, the little humanoid robot from softbank robotics. https://www.ald.softbankrobotics.com/en/cool-robots/ nao.

[10] Allison Sauppé and Bilge Mutlu. Effective task training strategies for human and robot instructors. Autonomous Robots, 39(3):313–329, 2015.

[11] Brian Scassellati, Henny Admoni, and Maja Matarić. Robots for use in autism research. Annual review of biomedical engineering, 14:275–294, 2012.

[12] [12] Sebastian Schneider, Michael Goerlich, and Franz Kummert. A framework for designing socially assistive robot interactions. Cognitive Systems Research, 43:301 – 312, 2017.

[13] Zhuoyu Shen and Yan Wu. Investigation of practical use of humanoid robots in elderly care centres. In Proceedings of the Fourth International Conference on Human Agent Interaction, HAI '16, pages 63–66, New York, NY, USA, 2016. ACM.

[14] Kenta Yamada and Jun Miura. Ambiguity-driven interaction in robot-to-human teaching. In Proceedings of the Fourth International Conference on Human Agent Interaction, pages 257–260. ACM, 2016.